\documentclass[sigconf]{acmart}
\AtBeginDocument{%
 }

\copyrightyear{2024}
\acmYear{2024}
\setcopyright{rightsretained}
\acmConference[WWW '24 Companion]{Companion Proceedings of the ACM Web Conference 2024}{May 13--17, 2024}{Singapore, Singapore}
\acmBooktitle{Companion Proceedings of the ACM Web Conference 2024 (WWW '24 Companion), May 13--17, 2024, Singapore, Singapore}\acmDOI{10.1145/3589335.3651459}
\acmISBN{979-8-4007-0172-6/24/05}
\usepackage{hyperref}
\usepackage{paralist}
\usepackage[inline]{enumitem}
\usepackage{xcolor}

\usepackage{subcaption}

\usepackage{multirow}

\usepackage[strict]{changepage}
\usepackage{framed}
\definecolor{formalshade}{rgb}{0.53,0.8,0.62}
\definecolor{darkblue}{rgb}{0.10,0.09,0.09}

\definecolor{spotgreen}{rgb}{0.03,0.48,0.18}

\usepackage{adjustbox}
\usepackage{array}

\newcolumntype{R}[2]{%
    >{\adjustbox{angle=#1,lap=\width-(#2)}\bgroup}%
    l%
    <{\egroup}%
}
\newcommand*\rot{\multicolumn{1}{R{0}{1em}}}

\newcommand\api[1]{{\fontfamily{pcr}\selectfont {\footnotesize #1}}}

\newcommand\feature[1]{{\footnotesize{\fontfamily{qhv}\selectfont #1}}}
\newcommand\res[2]{\small{$#1$}\tiny{\text{$\pm #2$}}}
\newcommand\bestres[2]{\small{$\mathbf{#1}$}\tiny{\text{$\mathbf{\pm #2}$}}}

\makeatletter
\gdef\@copyrightpermission{

   \href{https://creativecommons.org/licenses/by/4.0/}{This work is licensed under a Creative Commons Attribution International 4.0 License.}

  \vspace{5pt}
}
\makeatother

\begin{document}

\title{``All of Me'': Mining Users' Attributes from their Public Spotify Playlists}




\author{Pier Paolo Tricomi}
\orcid{0000-0003-1600-835X}
\affiliation{%
 \department{Department of Mathematics}
 \institution{University of Padova} 
 \city{Padova}
 \country{Italy}}
 \additionalaffiliation{
  \institution{Spritz Matter Srl}
 \city{Padova}
 \country{Italy}}
\email{pierpaolo.tricomi@phd.unipd.it}

\author{Luca Pajola}
\orcid{0000-0002-6749-6608}
\affiliation{%
 \institution{Spritz Matter Srl} 
 \city{Padova}
 \country{Italy}}
 \additionalaffiliation{%
 \institution{University of Padova}
 \city{Padova}
 \country{Italy}}
\email{luca.pajola@spritzmatter.com}

\author{Luca Pasa}
\orcid{0000-0002-3023-3046}
\affiliation{%
 \department{Department of Mathematics}
 \institution{University of Padova}
 \city{Padova}
 \country{Italy}}
\email{luca.pasa@unipd.it}

\author{Mauro Conti}
\orcid{0000-0002-3612-1934}
\affiliation{%
 \department{Department of Mathematics}
 \institution{University of Padova}
 \city{Padova}
 \country{Italy}}
\email{mauro.conti@unipd.it}





\renewcommand{\shortauthors}{Pier Paolo Tricomi, Luca Pajola, Luca Pasa, \& Mauro Conti}

\begin{abstract}
In the age of digital music streaming, playlists on platforms like Spotify have become an integral part of individuals' musical experiences. 
People create and publicly share their own playlists to express their musical tastes, promote the discovery of their favorite artists, and foster social connections.
In this work, we aim to address the question: can we infer users' private attributes from their public Spotify playlists?
To this end, we conducted an online survey involving 739 Spotify users, resulting in a dataset of 10,286 publicly shared playlists comprising over 200,000 unique songs and 55,000 artists. 
Then, we utilize statistical analyses and machine learning algorithms to build accurate predictive models for users' attributes.
\end{abstract}


\maketitle

\section{Introduction}
\label{sec.introduction}

The diffusion of music streaming services has revolutionized the way we consume and interact with music. With its user-friendly interface and vast music library, Spotify has become incredibly popular, reaching almost 500 Million users in 2022~\cite{spotifysurv}. 
Among its functionalities, Spotify allows users to create and \textit{publicly} share their playlists.
Previous studies demonstrated that people's music preferences can be linked to their private attributes, such as their personality traits~\cite{greenberg2016song}. If the public playlists on Spotify, representing users' music preferences, have the potential to disclose their private information,  the implications would be far-reaching. On one hand, the platform could deliver highly personalized content and empower content curators to craft playlists that resonate with distinct demographics and moods. However, on the other hand, the inference feasibility might pose privacy threats to platforms' users. 
The current landscape places significant emphasis on user profiling, evident in regulations like the EU AI Act, which will prohibit the training of models for biometric identification and categorization of individuals~\cite{aiact}. Consequently, it would be crucial for companies to proactively assess the risks associated with publicly releasing data (e.g., through APIs), and take appropriate measures.

\paragraph{Contribution}
We investigate the relationship between Spotify users’ attributes and their public playlists. Unlike earlier studies, we adopt a comprehensive approach, scrutinizing a wide range of 16 attributes spanning Demographic, Habits, and Personality related information. Our dataset encompasses over 10,000 playlists shared by 739 users, spanning over 200,000 songs and 55,000 artists. Our rigorous statistical analyses reveal the existence of a link between users' playlists and their personal attributes, while our Machine Learning (ML) models can predict them with appreciable accuracy (e.g., gender with more than 70\%).
We summarize our contributions:
\begin{itemize}
    \item We (ethically) collected a dataset of over 10,000 playlists shared by 739 Spotify users.
    \item We assess connections between users' public playlists and 16 private attributes, including unexplored ones.
    \item We showcase the feasibility of inferring users' attributes from their public playlists through a comprehensive ML testbed.
\end{itemize}

\paragraph{Transparency} The source code and the anonymized dataset are available at \url{https://github.com/pierz95/SpotifyAttributes}.

\begin{table*}[!t]
\caption{Target Attributes, their explanation, and distribution at user and playlist levels.}
\vspace{-1em}
\label{tab:distribution}
\footnotesize
\resizebox{\textwidth}{!}{%
\begin{tabular}{ccccc} \toprule
& \textbf{Target} &
  \textbf{Explanation} &
  \textbf{Distribution (User Level)} &
  \textbf{Distribution (Playlist Level)} \\ \midrule
& \textbf{Gender} &
  gender identity &
  Female (28\%), Male (68\%), Other (4\%) &
  Female (30\%), Male (66\%), Other (4\%) \\
& \textbf{Age} &
  current age &
  13-17 (15\%), 18-24 (45\%), 25-30 (29\%), 31+ (11\%) &
  13-17 (9\%), 18-24 (39\%), 25-30 (33\%), 31+ (19\%) \\
& \textbf{Country} &
  country of residence &
  \begin{tabular}[c]{@{}l@{}}US (27\%), IT (10\%), UK (7\%), CA (6\%), DE (5\%), \\ PH (3\%), AU (3\%), BR (3\%), IN (3\%), Other (33\%)\end{tabular} &
  \begin{tabular}[c]{@{}l@{}}US (32\%), IT (8\%), UK (16\%), CA (10\%), DE (3\%), \\ PH (2\%), AU (2\%), BR (3\%), IN (2\%), Other (22\%)\end{tabular} \\
& \textbf{Relashionship} &
  Whether a user is in a relationship &
  Yes (33\%), No (67\%) &
  Yes (45\%), No (55\%) \\
& \textbf{Live Alone} &
  Whether a user lives alone &
  Yes (14\%), No (86\%) &
  Yes (12\%), No (88\%) \\
& \textbf{Occupation} &
  Whether a user is employed &
  Yes (48\%), No (52\%) &
  Yes (61\%), No (39\%) \\
\parbox[t]{2mm}{\multirow{-7}{*}{\rotatebox[origin=c]{90}{\textbf{Demographic}}}} & \textbf{Economic} &
  Economic status (self reported) &
  Low (25\%), Medium (52\%), High (23\%) &
  Low (25\%), Medium (46\%), High (29\%) \\\midrule
& \textbf{Sport} &
  Whether a user does sport &
  Regularly (34\%), Occasionally (35\%), No (31\%) &
  Regularly (40\%), Occasionally (32\%), No (27\%) \\
& \textbf{Smoke} &
  Whether a user smokes &
  Yes (20\%), No (80\%) &
  Yes (20\%), No (80\%) \\
& \textbf{Alcohol} &
  Whether a user drinks alcohol &
  Yes  (54\%), No (46\%) &
  Yes  (66\%), No (34\%) \\
\parbox[t]{2mm}{\multirow{-4}{*}{\rotatebox[origin=c]{90}{\textbf{Habits}}}} & \textbf{Premium} &
  Whether a user has a Spotify Premium subscription &
  Yes (76\%), No (24\%) &
  Yes (88\%), No (12\%) \\\midrule
& \textbf{Openness} &
  Open-minded/curious (high) vs. consistent/cautious (low) &
  Low (7\%), Medium (46\%), High (47\%) &
  Low (3\%), Medium (42\%), High (55\%) \\
& \textbf{Conscientiousness} &
  Efficient/organized (high) vs extravagant/careless (low) &
  Low (20\%), Medium (62\%), High (18\%) &
  Low (23\%), Medium (56\%), High (21\%) \\
& \textbf{Extraversion} &
  Outgoing/energetic (high) vs. solitary/reserved (low) &
  Low (43\%), Medium (44\%), High (13\%) &
  Low (38\%), Medium (44\%), High (18\%) \\
& \textbf{Agreeableness} &
  Friendly/compassionate (high) vs. critical/rational (low) &
  Low (10\%), Medium (55\%), High (35\%) &
  Low (9\%), Medium (60\%), High (31\%) \\
\parbox[t]{2mm}{\multirow{-5}{*}{\rotatebox[origin=c]{90}{\textbf{Personality}}}}  & \textbf{Neuroticism} &
  Sensitive/nervous (high) vs. resilient/confident (low) &
  Low (23\%), Medium (41\%), High (36\%) &
  Low (23\%), Medium (43\%), High (34\%) \\ \bottomrule
\end{tabular}
}
\end{table*}
\section{Related Works}
\label{sec:related}
Several studies have attempted to link musical choices and personal attributes, especially personality traits~\cite{rentfrow2003re,rentfrow2012role,greenberg2016song}, while a minority concerned demographic factors (e.g., age~\cite{north2008social}) and cultural background~\cite{sloboda2001emotions}. Some previous works analyzed the correlations between users’ attributes and music data on music streaming service platforms, e.g., gender, age, and nationality on Last.fm~\cite{liu2012inferring,krismayer2019predicting}. 
Recently, a Spotify team investigated the interplay between personality traits and music listening habits~\cite{anderson2021just}, utilizing listening history, demographic data, and App Usage information (not publicly accessible) as model inputs. Our work distinguishes itself by i) concentrating on playlist information, a relatively unexplored aspect, ii) conducting a comprehensive analysis on 16 diverse attributes (demographics, habits, and personality), introducing novel elements (e.g., occupation, smoking habits), and iii) centering exclusively on publicly accessible data through the official Spotify API. 
Regarding personal attributes and public data, our paper acknowledges the cybersecurity research area exploring Attribute Inference Attacks (AIAs)~\cite{tricomi2023attribute}, clarifying our intent to understand connections between user preferences and behavior rather than pursuing AIAs. However, we highlight the privacy implications of uncovering such connections, emphasizing the need for awareness in handling user data on platforms like Spotify.

\section{Dataset}\label{sec:dataset}

\begin{figure*}[!ht]
\centering
\begin{subfigure}{.33\textwidth}
    \centering
    \includegraphics[width=.90\linewidth]{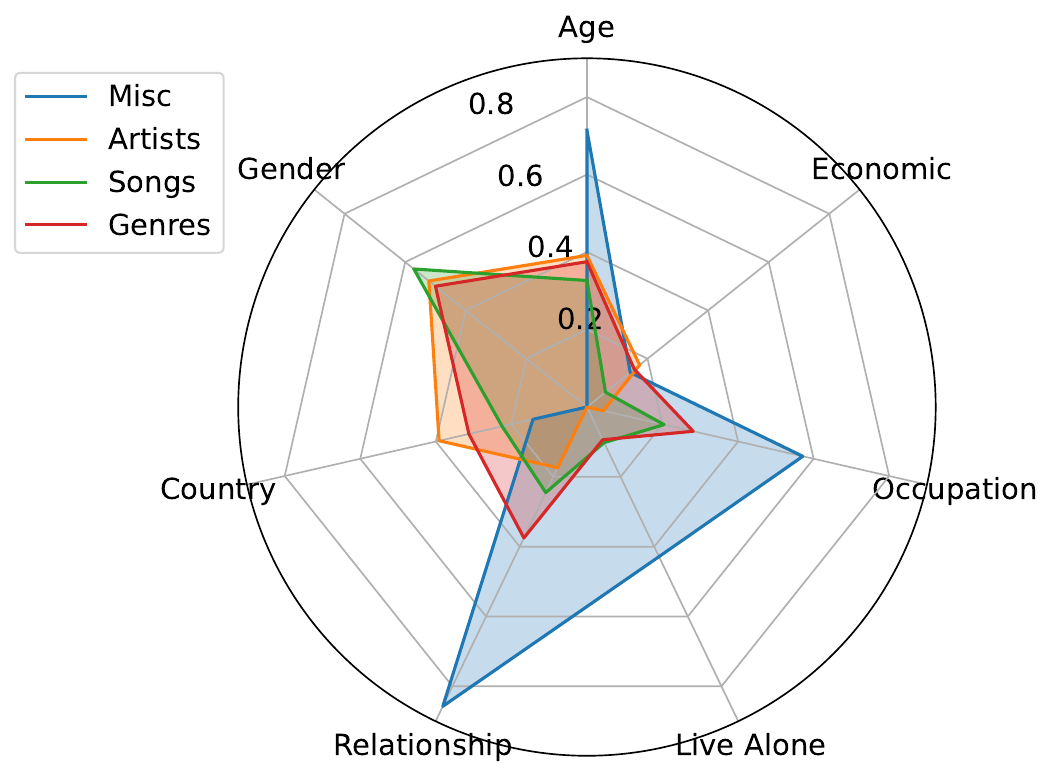}  
    \caption{Demographic}
    \label{SUBFIGURE LABEL 1}
\end{subfigure}
\begin{subfigure}{.33\textwidth}
    \centering
    \includegraphics[width=.90\linewidth]{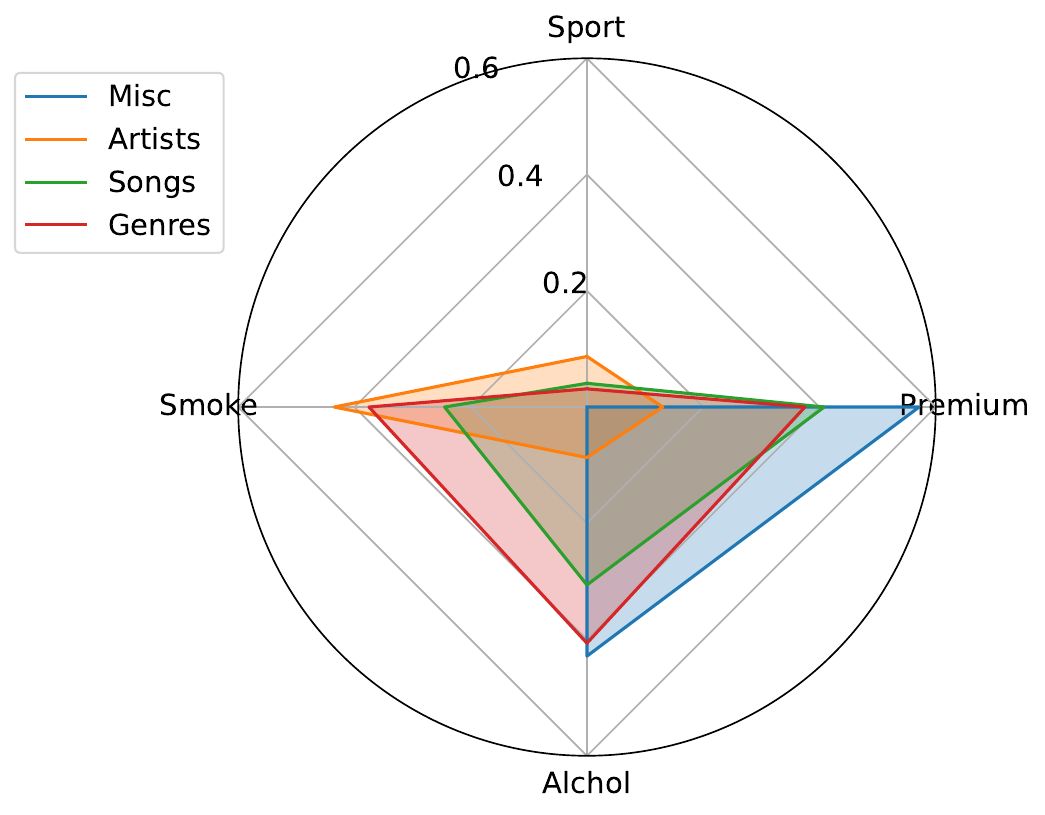}  
    \caption{Habits}
    \label{SUBFIGURE LABEL 2}
\end{subfigure}
\begin{subfigure}{.33\textwidth}
    \centering
    \includegraphics[width=.90\linewidth]{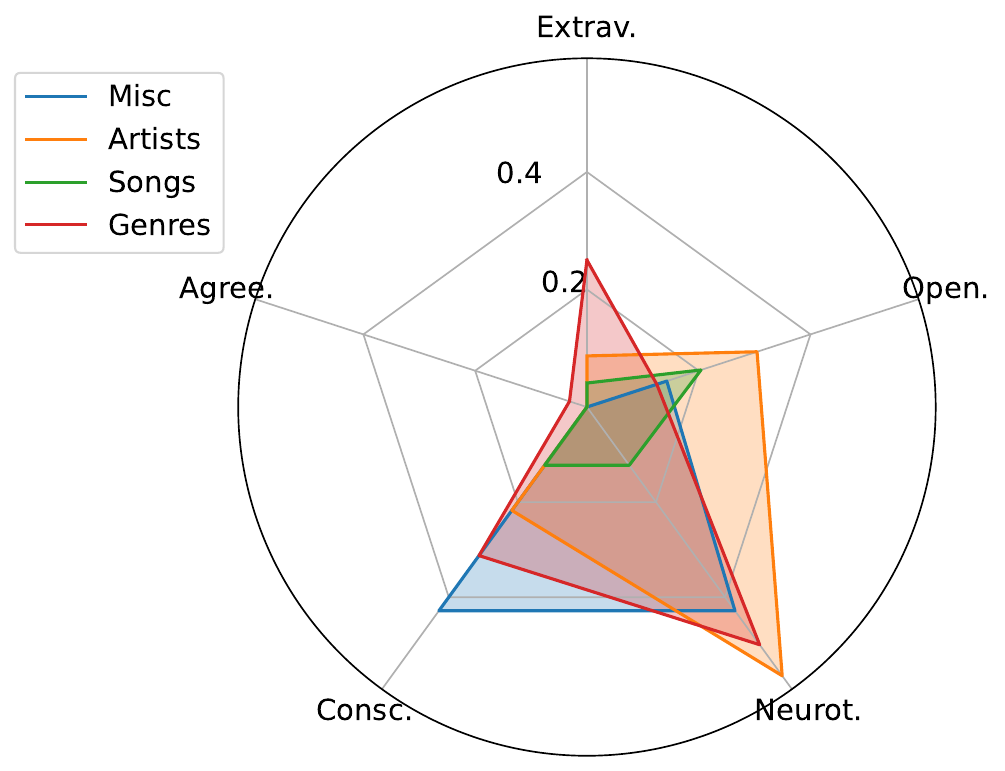}  
    \caption{Personality}
    \label{SUBFIGURE LABEL 3}
\end{subfigure}
\vspace{-2em}
\caption{Ratio of feature families that exhibit statistically significant distinctions between attribute groups.} 
\label{fig:spider}
\end{figure*}

\paragraph{The Survey} 
For our dataset, we recruited participants via an online survey. We asked them to provide their Spotify IDs necessary to access their playlists, and additional self-provided information to serve as attributes for our analyses. In total, we gathered 16 attributes divided into demographics, habits, and personality domains. We utilized the 10 short personality questions to retrieve OCEAN personality traits~\cite{rammstedt2007measuring}.
Table~\ref{tab:distribution} reports a short description of the attributes. 
The survey incorporated several attention checks to filter out inconsistent or unreliable responses. The survey was active from May to September 2022, and distributed primarily through social networking platforms, notably Reddit, Facebook, and Telegram. Our survey was strategically shared within popular Spotify and music-oriented groups, including \textit{/r/Spotify/}, \textit{r/Music}, and \textit{Facebook Spotify Music}. 
Users explicitly permitted us to utilize their data for this study by participating in the survey. We also offered various points of contact to request data removal, and anonymized their data, eliminating any Personally Identifiable Information (PII).

\paragraph{Survey results and validation} We received 739 valid responses from 76 different countries, age range 13--55.\footnote{Participants under the age of 13 were excluded to adhere to European regulations, as further detailed in the Ethical Consideration section.} Table~\ref{tab:distribution} reports the user and playlist distributions of our targets, which can differ since users can have an arbitrary number of playlists.  
Similar to previous works~\cite{tricomi2023attribute,bunian2017modeling}, we grouped \feature{age} in bins of interest (e.g., underage), and personality traits into low, middle, and high scores ([0, 33.3), [33.3, 66.6), [66.6, 100]). 
Our distributions align to global Spotify statistics\cite{spotifysurv, spotifycountr}. 
Our number of Spotify users is above the minimum sample size~\cite{kotrlik2001organizational} of 384 required to draw statistically significant results, with a confidence level of 95\%, a margin of error of 5\%, a population proportion of 50\%, and a population size of 500 millions.

\textbf{Remark.} Our primary objective is to establish the existence of a connection between users' playlists and personal attributes. However, given the limited dataset and possible biases, this relationship may exhibit variations in a broader and more representative context (e.g., wider sample size, finer attributes granularity). Nonetheless, all our analyses are rigorously validated through statistical tests, ensuring the robustness and reliability of our findings.


\paragraph{Features}\label{ssec:dataset_features}
To create our playlist dataset, we harnessed multiple Spotify APIs\footnote{https://developer.spotify.com/}. We collected features regarding:
\begin{itemize}
    \item \textbf{Songs:} for each song in a playlist, we accessed the \api{tracks} and \api{audio-features} APIs to retrieve generic songs' information (e.g., release year, popularity) and audio features calculated by Spotify (e.g., danceability, acousticness). 
     \item \textbf{Artists:} through \api{artists} API, we extracted their popularity and number of followers. We also considered the (unique) number of (popular and unpopular) artists in a playlist and the artist diversity (measured through the Simpson Index~\cite{keylock2005simpson}).
    \item \textbf{Genres:} through \api{artists} API, we attributed to each song the genre of its artist(s). Then, we calculated the proportion of songs within the playlist that fell under specific genres, encompassing 30 popular genres (e.g., rock, pop, indie, metal). 
    \item \textbf{Miscellaneous:} for each playlist, we included the total number of songs, the count of followers, the diversity in terms of the albums from which the songs originated, and a record of the years when the songs were added to the playlist.
\end{itemize}

Then, we consolidated this information to craft a unified representation consisting of 111 features for each playlist, serving as the foundation for our experiments. For numeric features, we calculated the mean, standard deviation, min and max values as aggregation methods. 
The playlist was associated with the attributes of the user who created it (e.g., age, gender). 
The final playlist dataset comprehends 10,286 public playlists made by 739 users.
A detailed description of our features and dataset is available in our repository.

\section{Statistical Analyses}\label{sec:anal}


This section explores the relationships between users' personal attributes and their Spotify public playlists, based on our feature families. For instance, are artists-related features connected with demographic attributes? Are genre preferences connected with habits such as drinking alcohol or smoking? To unravel these connections comprehensively, we undertake a thorough examination of all possible combinations of attributes (Demographic, Habits, and Personality) and feature families (Misc, Artists, Songs, Genres).

\paragraph{Methodology}
To assess relationships between attributes and features, we utilized two widely recognized statistical tests: the unpaired Student t-test and ANOVA. These tests assess whether there are statistically significant differences between the means of two or more groups, determining whether any of these groups differ from each other in terms of a dependent variable. In our case, we treated our features, individually, as the dependent variable, with attributes serving as the distinct groups for assessment. Consequently, we conducted a battery of tests, subjecting each variable to examination against every attribute. Specifically, we used the Student t-test for attributes with two classes (e.g., Relationship) and ANOVA for others. The significance of these tests is established when the associated $p$-value falls below a predetermined threshold, typically set at 0.05. To mitigate the potential influence of users with a larger number of playlists, we performed these tests at the user level, where we aggregated all playlist information for each individual, ensuring a balanced and equitable analysis.

\paragraph{Results}
Figure~\ref{fig:spider} reports the results of our analysis, categorized by attribute types and feature families. These graphs illustrate the proportion of features within a family that contribute to statistically significant differences between the classes of the target attribute ($p\!<\!0.05$). We notice that different groups of features help tell apart different attributes.
For demographic attributes, the classes of \feature{Age}, \feature{Relationship}, and \feature{Occupation} exhibit significant distinctions primarily based on Misc features, while \feature{Gender} showcases the least variation. Conversely, \feature{Gender} classes differ from one another in Artists, Songs, and Genres features, a pattern shared by \feature{Age}, while \feature{Economic} and \feature{Live Alone} do not demonstrate strong relationships.
For Habits attributes, \feature{Alcohol}, \feature{Smoke}, and \feature{Spotify Premium} reveal notable differences across all features, while \feature{Sport} exhibits relatively milder associations. A parallel pattern emerges in the context of Personality Attributes, where \feature{Agreeableness} and \feature{Extraversion} exhibit fewer disparities in our features, while \feature{Neuroticism}, \feature{Openness}, and \feature{Conscientiousness} reveal the most pronounced distinctions. To sum up, playlists' information and user attributes are statistically related.

\begin{figure}[!h]
    \centering
    \includegraphics[width=0.80\linewidth]{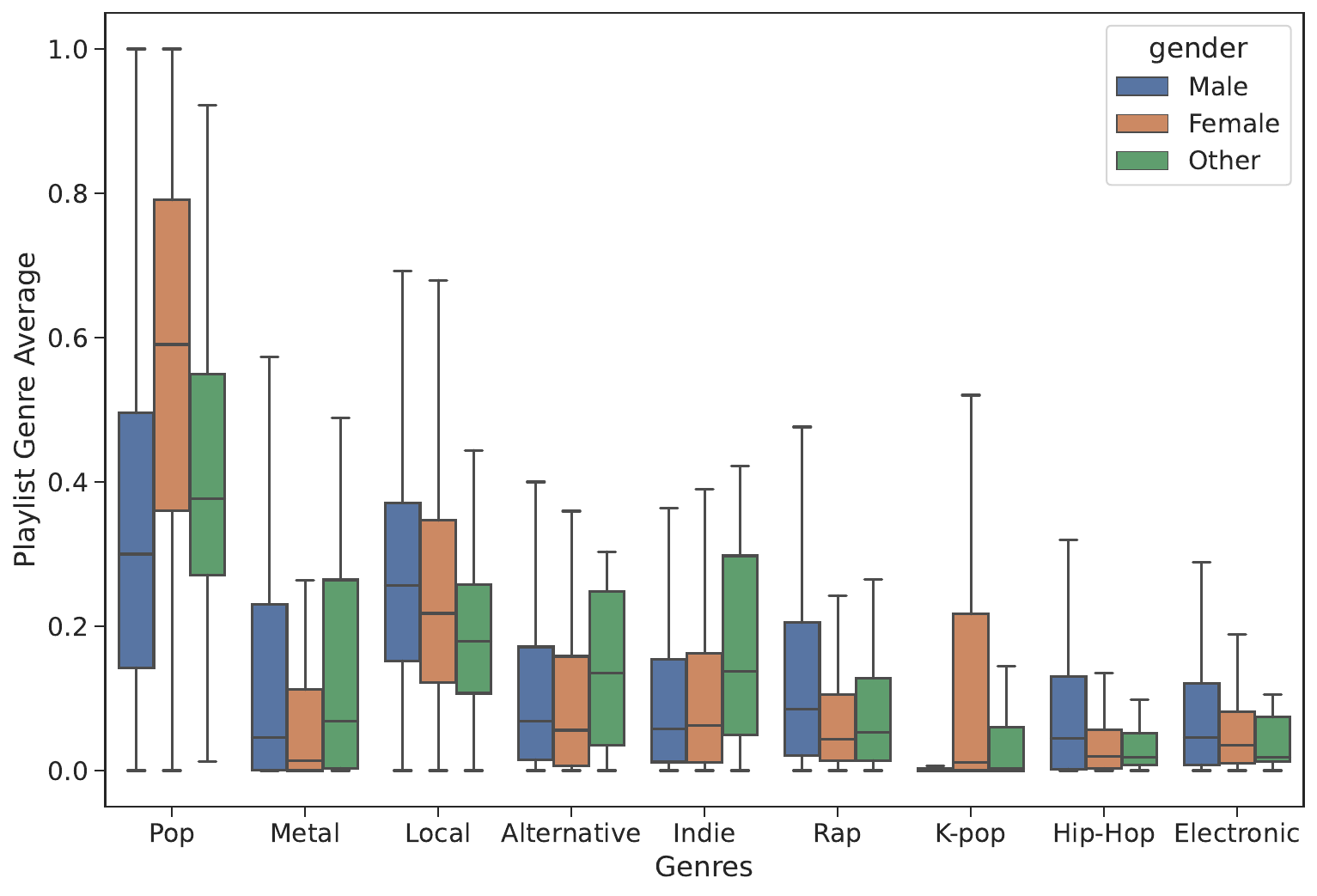}
    \vspace{-1em}
    \caption{Genres distributions per Gender attribute.}
    \vspace{-1em}
    \label{fig:genres}
\end{figure}

We delved into the distribution of \feature{Gender} for Genre features as an example to understand better these analyses (Figure~\ref{fig:genres}). The following p-values, calculated through ANOVA, indicate statistically significant differences. Males favorite genres are local ($p\!<\!0.05$), rap, hip-hop, and electric ($p\!<\!0.001$). Almost no male listen to K-pop ($p\!<\!0.001$). 
Females, instead, tend to prefer pop and k-pop genres ($p\!<\!0.001$), while showing less interest in metal ($p\!<\!0.05$). 
Interestingly, non-binary individuals often exhibit preferences that fall between Males and Females, with a particular inclination toward alternative and indie genres ($p\!<\!0.001$). 

\section{Predicting Users' attributes}\label{sec:class}

\begin{table*}[]
    \centering
    
    \caption{Classification Results (best in red). Smoke and S. Premium best models are weakly statistically significantly better than RG ($p$-value<0.10). All the others, except for Live A., are statistically significantly better than RG ($p$-value<0.05) }
    \vspace{-1em}
    \label{tab:class-demo-all}
    \resizebox{\linewidth}{!}{%
    \begin{tabular}{c|ccccccc|cccc|ccccc}
        
        \toprule
        & \multicolumn{7}{c|}{\textbf{Demographic}} & \multicolumn{4}{c|}{\textbf{Habits}} & \multicolumn{5}{c}{\textbf{Personality}} \\
        {\textbf{Model}} &  \textbf{Age} &  \textbf{Country} &         \textbf{Econ.} &           \textbf{Gender} &        \textbf{Live A.} &   \textbf{Relat.} &        \textbf{Occup.} &           \rot{\textbf{Alchol}} &            \rot{\textbf{Smoke}} & \rot{\textbf{Sport}} &  \textbf{S. Premium}  &         \textbf{Open.} & \textbf{Consc.} &     \textbf{Extrav.} &    \textbf{Agree.} &      \textbf{Neurot.} \\
        \midrule
        RG        &  \res{33.6}{3.4} &  \res{13.7}{3.3} &  \res{35.6}{0.0} &  \res{53.6}{2.1} &  {\color{red} \bestres{80.2}{0.0}} &  \res{50.0}{4.3} &   \res{47.7}{4.1} &  \res{51.6}{4.2} &  \res{70.7}{0.0} &  \res{22.4}{6.4} &  \res{65.2}{0.0} &  \res{31.2}{1.9} &   \res{47.7}{0.0} &  \res{34.1}{4.3} &  \res{38.7}{0.0} &  \res{23.8}{1.3}\\\hline
        LR       &  \res{40.1}{5.2} &  \res{27.6}{2.3} &  \res{38.2}{1.6} &  \res{67.9}{1.9} &  \res{80.2}{0.0} &  \color{red}\bestres{63.0}{2.8} &  \res{60.4}{10.1}  &  \res{54.9}{4.1} &  \res{68.8}{7.3} &  \res{33.7}{3.8} &  \res{65.4}{0.8}  &  \res{40.7}{4.9} &   \res{49.3}{1.2} &  \color{red}\bestres{42.9}{2.6} &  \res{39.8}{3.8} &  \res{36.5}{4.5}  \\
        DT &  \res{40.4}{3.6} &  \res{24.1}{2.4} &  \color{red}\bestres{40.3}{2.4} &  \res{68.3}{1.2} &  \res{80.2}{0.0} &  \res{58.9}{5.8} &   \res{57.0}{4.7} & \res{53.6}{6.0} &  \color{red}\bestres{72.0}{1.8} &  \res{31.4}{4.2} &  \res{65.0}{1.2} &  \res{43.1}{4.9} &   \res{47.8}{0.8} &  \res{38.3}{3.8} &  \res{43.6}{1.3} &  \res{39.6}{6.1}  \\
        RF &  \color{red}\bestres{42.2}{6.5} &  \res{26.8}{2.0} &  \res{38.5}{3.6} &  \res{67.8}{1.0} &  \res{80.2}{0.0} &  \res{63.0}{2.9} &   \res{60.9}{3.5} & \res{54.9}{5.0} &  \res{71.7}{1.5} &  \res{32.4}{1.9} &  \color{red}\bestres{66.9}{1.9}  &  \res{45.5}{6.1} &   \res{48.9}{1.4} &  \res{42.8}{4.8} &  \res{43.7}{2.7} &  {\color{red} \bestres{40.2}{4.1}}   \\
        KNN          &  \res{36.7}{3.0} &  \res{27.6}{2.3} &  \res{38.7}{2.7} &  \color{red}\bestres{70.8}{2.2} &  \res{80.1}{0.2} &  \res{61.6}{2.1} &   \res{59.5}{6.0} &  \res{53.5}{3.4} &  \res{70.7}{1.7} &  {\color{red}\bestres{33.8}{2.9}} &  \res{66.9}{1.1} &  {\color{red}\bestres{46.6}{2.6}} &   \res{48.6}{1.5} &  \res{42.2}{4.6} &  {\color{red}\bestres{46.5}{5.7}} &  \res{34.4}{2.1}  \\
        MLP          &  \res{40.8}{2.8} &  \color{red}\bestres{31.5}{2.7} &  \res{38.9}{3.4} &  \res{68.0}{3.1} &  \res{80.2}{0.0} &  \res{62.3}{4.1} &   {\color{red}\bestres{63.7}{5.7}} &  \color{red}\bestres{55.6}{2.3} &  \res{71.7}{1.8} &  \res{32.2}{3.0} &  \res{66.2}{1.8} &  \res{45.2}{2.9} &  {\color{red} \bestres{50.4}{2.1}} &  \res{42.1}{4.1} &  \res{43.5}{3.0} &  \res{39.3}{3.9} \\
        \bottomrule
        \end{tabular}
        
    }
\end{table*}

\paragraph{Dataset} 
We split the playlists' dataset into training, validation, and testing sets in a stratified fashion, with 70\% - 10\% - 20\% ratios, respectively. 
The split is based on the users, i.e., all the playlists of a given user belong to the same partition. Indeed, in a realistic scenario, it is unlikely to have the same user in both training and testing sets, as we would already know their attribute.

\paragraph{Models}
Given the user's playlists, we want to predict the user's class for each attribute (e.g., Low, Medium, or High Economic). 
As models, we tested Logistic Regression (LR), Decision Tree (DT), Random Forest (RF), K-Nearest-Neighbours (KNN), and MultiLayer Perceptron (MLP).
All of these models assume a fixed input size, while in our case, the number of playlists varies user by user. 
Combining playlists into a fixed-size input vector is not ideal because it is hard to decide how to pad and arrange them without adding bias.
Therefore, the models process one playlist at a time, producing a classification (probability) for each of them. For each user, we then combine the output probabilities of all their playlists and calculate the average classifications. 
We employed a comprehensive model selection strategy to choose hyperparameters (more details are available in our repository), ensuring a fair and reproducible comparison of model performances. We adopted a grid search approach, evaluating the models using a weighted f1-score. We also added a stratified random guess (RG)\footnote{https://scikit-learn.org/stable/modules/generated/sklearn.dummy.DummyClassifier.html} as a baseline, and repeated each experiment five times, varying the users of the three sets.

\paragraph{Results}
Table~\ref{tab:class-demo-all} shows the results.
We outperform the baseline (RG) in 15/16 cases, indicating the feasibility of inference. All these best models are statistically significantly better than RG ($p$-value<0.05, Unpaired Student t-test), except for Smoke and S. Premium attributes, which are weakly statistically significantly better ($p$-value<0.1). There is no universal best algorithm, suggesting tailored experiments may be needed for each attribute. The only exception is the \feature{Live Alone} attribute, which cannot be inferred with our extracted features. Overall, it is possible to infer a broad range of user attributes from Spotify public playlists, with the best models achieving at least 10 percentage points higher accuracy than random guessing for attributes like \feature{Country}, \feature{Gender}, \feature{Relationship}, \feature{Occupation}, \feature{Sport}, \feature{Openness}, and \feature{Neuroticism}.

\section{Conclusion and Future Works}
\label{sec:concl}
This study has revealed a substantial connection between Spotify playlist information and user attributes, with specific attributes, such as \feature{Age} and \feature{Gender}, exhibiting stronger correlations than others like \feature{Living Alone}. ML achieved appreciable results in harnessing these features for attribute prediction, which raises awareness regarding Spotify users' privacy. In the future, we plan to extend our statistical analysis to gain more insight regarding the link between playlists and private attributes, and employ more sophisticated algorithms to increase predictive performances.


\subsection*{Ethical Considerations}
\label{ssec:ethical}
Our institutions do not mandate formal IRB approval for the experiments detailed in this study. However, we conducted our survey and evaluations in strict accordance with the ethical guidelines outlined in the Menlo report~\cite{bailey2012menlo}. Participants were explicitly told their responses would be used for research, and our questionnaire did not ask for sensitive or personally identifiable information like names or addresses. We provided participants with an easily accessible email contact for requesting the removal of their entries from our dataset. We reported only aggregated results to avoid risks of de-anonymization. Given we are based in Europe, we diligently adhered to GDPR regulations. Furthermore, all underage participants were situated in regions where their involvement in research surveys did not necessitate explicit parental consent~\cite{fra2020child}.

\bibliographystyle{ACM-Reference-Format}
\bibliography{bibliography}



\end{document}